\documentstyle[12pt]{article}
\input psfig.sty
\def\nn{\noindent}

\def\Re{{\cal R \mskip-4mu \lower.1ex \hbox{\it e}\,}}
\def\Im{{\cal I \mskip-5mu \lower.1ex \hbox{\it m}\,}}
\def\ie{{\it i.e.}}

\def\etal{{\it et al.}}
\def\ibid{{\it ibid}.}
\def\sub#1{_{\lower.25ex\hbox{$\scriptstyle#1$}}}

\def\to{\rightarrow}

\def\subw{_{\rm w}}
\def\mh{\ifmmode m\sbl H \else $m\sbl H$\fi}
\def\mch{\ifmmode m_{H^\pm} \else $m_{H^\pm}$\fi}
\def\mt{\ifmmode m_t\else $m_t$\fi}
\def\mc{\ifmmode m_c\else $m_c$\fi}
\def\mz{\ifmmode M_Z\else $M_Z$\fi}
\def\mw{\ifmmode M_W\else $M_W$\fi}
\def\mws{\ifmmode M_W^2 \else $M_W^2$\fi}
\def\mhs{\ifmmode m_H^2 \else $m_H^2$\fi}   
\def\mzs{\ifmmode M_Z^2 \else $M_Z^2$\fi}
\def\mts{\ifmmode m_t^2 \else $m_t^2$\fi}
\def\mcs{\ifmmode m_c^2 \else $m_c^2$\fi}
\def\mchs{\ifmmode m_{H^\pm}^2 \else $m_{H^\pm}^2$\fi}
\def\ztwo{\ifmmode Z_2\else $Z_2$\fi}
\def\zone{\ifmmode Z_1\else $Z_1$\fi}
\def\mtwo{\ifmmode M_2\else $M_2$\fi}
\def\mone{\ifmmode M_1\else $M_1$\fi}
\def\tb{\ifmmode \tan\beta \else $\tan\beta$\fi}
\def\xw{\ifmmode x\subw\else $x\subw$\fi}
\def\ch{\ifmmode H^\pm \else $H^\pm$\fi}
\def\lum{\ifmmode {\cal L}\else ${\cal L}$\fi}
\def\inpb{\ifmmode {\rm pb}^{-1}\else ${\rm pb}^{-1}$\fi}
\def\infb{\ifmmode {\rm fb}^{-1}\else ${\rm fb}^{-1}$\fi}
\def\epem{\ifmmode e^+e^-\else $e^+e^-$\fi}
\def\ppb{\ifmmode \bar pp\else $\bar pp$\fi}
\def\bsg{\ifmmode B\to X_s\gamma\else $B\to X_s\gamma$\fi}
\def\bsll{\ifmmode B\to X_s\ell^+\ell^-\else $B\to X_s\ell^+\ell^-$\fi}
\def\bstt{\ifmmode B\to X_s\tau^+\tau^-\else $B\to X_s\tau^+\tau^-$\fi}

\newskip\zatskip \zatskip=0pt plus0pt minus0pt
\def\matth{\mathsurround=0pt}
\def\lsim{\mathrel{\mathpalette\atversim<}}

\def\be{\begin{equation}}
\def\ee{\end{equation}}
\def\bea{\begin{eqnarray}}
\def\eea{\end{eqnarray}}
\def\atversim#1#2{\lower0.7ex\vbox{\baselineskip\zatskip\lineskip\zatskip
  \lineskiplimit 0pt\ialign{$\matth#1\hfil##\hfil$\crcr#2\crcr\sim\crcr}}}

\renewcommand{\thefootnote}{\fnsymbol{footnote}}

\begin{document} \begin{titlepage} 
\rightline{\vbox{\halign{&#\hfil\cr
&MADPH-98-1090\cr
&SLAC-PUB-7981\cr
&November 1998\cr}}}
\begin{center}

{\Large\bf
Top-Charm Associated Production in
High Energy $e^+e^-$ Collisions}
\footnote{Work supported by the Department of 
Energy, Contract DE-AC03-76SF00515}
\medskip

\normalsize 
{\large Tao Han$^{(a)}$ and JoAnne L. Hewett$^{(b)}$ } \\
\vskip .3cm
$^{(a)}$Department of Physics, University of Wisconsin, Madison, WI 53706\\
$^{(b)}$Stanford Linear Accelerator Center,
Stanford CA 94309, USA\\
\vskip .3cm

\end{center}

\begin{abstract} 

The possibility of exploring the flavor changing neutral current 
$tcZ/tc\gamma$ couplings in the production vertex for the reaction 
$\epem\to t\bar c + \bar tc$ is examined. Using a model independent 
parameterization for the effective Lagrangian to describe the most general 
three-point interactions, production cross sections are found to be 
relatively
small at LEP II, but potentially sizeable at higher energy \epem\ colliders.  
The kinematic characteristics of the signal are studied and a set of cuts 
are devised for clean separation of the signal from background.  The
resulting sensitivity to anomalous flavor changing couplings at LEP II 
with an integrated luminosity of $4\times 500$ pb$^{-1}$ is found to be 
comparable to their present indirect constraints from loop processes, while at
higher energy colliders with $0.5-1$ TeV center-of-mass energy and
$50-200$ fb$^{-1}$ luminosity, one expects to reach a sensitivity
at or below the percentage level.

\end{abstract}

\renewcommand{\thefootnote}{\arabic{footnote}} \end{titlepage} 


\section{Introduction}  

It is often stated that the large value of the top-quark mass opens the
possibility that it plays a special role in particle physics.  
Indeed, the properties of the top-quark could reveal information on the
nature of electroweak symmetry breaking, address questions in flavor 
physics,
or provide special insight to new interactions originating at a higher 
scale.
One consequence of its large mass is that top decays rapidly via $t\to W+b$,
before the characteristic time for hadron formation and hence
top-flavored meson states do not form.  This results in a fundamentally
different phenomenology for top than for the lighter quark states and allows 
for the unique capability to determine the properties of the quark 
itself\cite{smtop}.
The precise determination of these properties may well reveal the existence
of physics beyond the Standard Model (SM)\cite{bsmtop}.

One possible manifestation of new interactions in the top-quark sector is
to alter its couplings to the gauge bosons.  Such anomalous couplings
would modify top production and decay at colliders\cite{topprod}, as well as
affecting loop-induced processes\cite{loops}.  The most widely studied
cases are the $t\bar tV$, with $V=\gamma\,, Z\,, g$, and $tbW$
three-point functions.  However, the flavor changing neutral current (FCNC)
interactions $t\bar cV\,, t\bar uV$ also offer an ideal place to 
search for new physics as they are very small in the SM\cite{ehs}. 
In this instance any positive observation of these transitions 
would unambiguously signal the presence of new physics.  
The FCNC vertices can be probed either
directly in top-quark decays, indirectly in loops, or via the production
vertex for top plus light quark associated production.  It is the latter
case which is studied here in the reaction 
\begin{equation}
\epem\to t\bar c+\bar tc.
\end{equation}
As will be discussed below, this mechanism offers some advantages due
to the ability to probe higher dimension operators at large momenta and
to striking kinematic signatures which are straightforward to detect in the 
clean environment of \epem\ collisions.

\section{Top FCNC Interactions}

Deviations from the SM for the flavor changing vertices can be described
by a linear effective Lagrangian which contains operators in an expansion 
series in powers of $1/\Lambda$, where $\Lambda$ is a high mass scale 
characteristic of the new interactions.  In this case 
the lowest dimension gauge invariant operators built from SM fields are 
dimension six and can be written as\cite{dimops,esma}
\bea
\label{dimsix}
{\cal L}_{eff} & = & {\alpha_B\over\Lambda^2} \bar Q_L\sigma^{\mu\nu}c_R
\tilde\Phi B_{\mu\nu}+{\alpha_W\over\Lambda^2} \bar Q_L\sigma^{\mu\nu}c_R
\tilde\Phi \tau_aW^a_{\mu\nu}\nonumber\\
& & +{\tilde\alpha_B\over\Lambda^2} \bar Q_L\sigma^{\mu\nu}i\gamma_5c_R
\tilde\Phi B_{\mu\nu}+{\tilde\alpha_W\over\Lambda^2} \bar Q_L\sigma^{\mu\nu}
i\gamma_5c_R\tilde\Phi \tau_aW^a_{\mu\nu}+h.c.\, ,
\eea
with $W^a_{\mu\nu}\,, B_{\mu\nu}$ being the field strength tensors for the
three  non-abelian fields of SU(2)$_L$ and the single abelian gauge field 
associated with U(1)$_Y$ respectively, $\tilde\Phi$ is the conjugate Higgs
field $\tilde\Phi=i\tau_2\Phi^*$, $\tau_a$ are the Pauli spin matrices, and
$Q_L$ represents the third generation left-handed quark doublet.  This 
effective Lagrangian is then added to that of the SM and after spontaneous 
symmetry breaking it induces the dimension five operators in
\bea
\label{effL}
{\cal L} & = & e\bar t {i\sigma_{\mu\nu}q^\nu\over m_t+m_c}
(\kappa_\gamma-i\tilde\kappa_\gamma\gamma_5)c A^\mu \\
& & +{g\over 2c_w}\bar t\left[ \gamma_\mu(v_Z-a_Z\gamma_5)+
{i\sigma_{\mu\nu}q^\nu\over 
m_t+m_c}(\kappa_Z-i\tilde\kappa_Z\gamma_5)\right]
cZ^\mu +h.c.\,,\nonumber
\eea
where $q$ labels the momentum of the gauge boson.
The dimension four terms can arise from anomalous contributions 
of terms like $(v^2/\Lambda^2)\bar\psi_i\gamma_\mu D^\mu \psi_j$,
where $v$ represents the vacuum expectation value of the SM Higgs field
and $D^\mu$ is the covariant derivative.  
The coefficients of the dimension 
five terms are related to those of the dimension six operators above by
\bea
{e\over m_t+m_c}~^(\tilde\kappa^)_\gamma & = & {\sqrt 2v\over
\Lambda^2}(c_w~^(\tilde\alpha_B^)+s_w~^(\tilde\alpha_W^))\,,\nonumber\\
{g\over 2c_w(m_t+m_c)}~^(\tilde\kappa^)_Z & = & {\sqrt 2v\over
\Lambda^2}(s_w~^(\tilde\alpha_B^)-c_w~^(\tilde\alpha_W^))\,.
\eea
In the flavor conserving case, $\kappa$ and $\tilde\kappa$ are the magnetic
and electric dipole moment form factors, respectively, of the fermion to the 
$\gamma$ and $Z$.  Note that the $\tilde\kappa$ terms are CP-violating.  It 
is
also possible to derive these operators from a non-linear effective 
Lagrangian
approach\cite{nonlin}, where the exact form of Eq. (\ref{effL}) can be
derived in the unitary gauge with prescribed relations\cite{esma} between 
the form factors and the parameters of the chiral expansion.  
In principle, the operator $e\bar tF(q^2)(\gamma_\mu q^2-q_\mu\not q)cA^\mu$
can also mediate FCNC interactions for non-zero values of $q^2$, \ie,
$F(q^2=0)=0$, but we do not consider this possibility here.  In the
following, we employ Eq. (\ref{effL}) as a model independent 
parameterization 
of the effects of new physics on the FCNC three-point function and 
assume that the form factors are static.

It is instructive to roughly estimate the relative sizes of the anomalous
couplings. It is sensible to assume that the couplings 
$\alpha,\tilde \alpha$ in Eq.~(\ref{dimsix}) are naturally of ${\cal O}(1)$.
We thus expect that $\kappa,\tilde\kappa$ as well as $a^{}_Z,v^{}_Z$
in Eq.~(\ref{effL}) to be of ${\cal O}(0.1)$. 
This estimate of course depends on the
normalization scale which we have conveniently chosen as 
$m_t$ in order to correspond to the traditional dipole moment form factor
definitions. If we took it to be scaled by $\Lambda=1$ TeV instead, 
then the $\kappa$'s would be roughly of order 1.

A convenient way to compare the sensitivity of various processes to these
anomalous form factors, as well as to evaluate their expected values in 
different models, is to relate them to the FCNC partial widths of the
top-quark.  In the case of the dimension four operators, this can be readily
computed and gives the branching fraction
\be
{\Gamma_4(t\to cZ)\over \Gamma(t\to bW)}
= {(m_t^2-M_Z^2)^2(m_t^2+2M_Z^2)\over (m_t^2-M_W^2)^2(m_t^2+2M_W^2)}
\ (v_Z^2+a_Z^2)\simeq (v_Z^2+a_Z^2)\,.
\ee
For the dimension five operators, the results in the on-shell case are
\be
{\Gamma_5(t\to cZ)\over \Gamma(t\to bW)}
= {(m_t^2-M_Z^2)^2(m_t^2+3/2M_Z^2)\over (m_t^2-M_W^2)^2(m_t^2+2M_W^2)}
\ {2M_Z^2\over  m_t^2}
\ (\kappa_Z^2+\tilde\kappa_Z^2)\simeq
\ 0.55 (\kappa_Z^2+\tilde\kappa_Z^2)\,.
\ee
and
\be
{\Gamma_5(t\to c\gamma)\over \Gamma(t\to bW)}
= {m_t^6\over (m_t^2-M_W^2)^2(m_t^2+2M_W^2)}
\ {4M_W^2\over  m_t^2}\sin^2\theta_W
\ (\kappa_\gamma^2+\tilde\kappa_\gamma^2)\simeq
\ 0.3 (\kappa_\gamma^2+\tilde\kappa_\gamma^2)\,.
\ee
As previously mentioned, the top FCNC branching fractions 
are unmeasureably small in the SM\cite{ehs}
with $B(t\to c\gamma\,, cZ)= 5.2\times 10^{-13}\,, 1.5\times 10^{-13}$,
respectively, for $m_t=175$ GeV.  Substantial enhancements of $4-5$
orders of magnitude can be obtained\cite{ehs,susy} in flavor conserving 
Two-Higgs-Doublet models and Supersymmetry, however the resulting branching 
ratios remain small being of order $10^{-9}-10^{-8}$ at the largest.  
Flavor changing
Multi-Higgs-Doublet models bring further enhancements\cite{laura} with
$B(t\to cV)\sim 10^{-6}-10^{-5}$ being possible.  However, models with 
singlet 
quarks\cite{berbarger} which contain tree-level FCNC, compositeness 
models\cite{comp}, or models of dynamical electroweak 
symmetry breaking\cite{sews}, which can all introduce effective flavor 
changing
couplings of ${\cal O}(\sqrt{m_tm_c}/v)$, can yield sizeable branching fractions
of $B(t\to cV)\lsim 1\%$.

Some models which induce FCNC are more naturally probed via top-charm
associated production than in flavor changing top-quark decays due to the 
large underlying mass scales and possibly large momentum transfer.  
An illustration of this, which is particularly
well suited to the reaction considered here, is that of Topcolor assisted
Technicolor\cite{birdman}.  Tree-level FCNC for the additional neutral gauge
boson present in this model are generated when the quark fields are rotated 
to the mass eigenstate basis.  The couplings of this $Z'$ are non-universal 
and stronger for the third generation, yielding potentially large $tcZ'$
interactions.  In addition, the production rate for this $Z'$, which is 
constrained\cite{tern} to be heavier than $\sim 1.5$ TeV,  is sizeable 
in high energy \epem\ collisions\cite{tgr}, and hence $\epem\to t\bar c+
\bar tc$ is the ideal place to search for this effect.
Another example is given by Multi-Higgs-Doublet models with tree-level FCNC.
In this case, s-channel Higgs exchange can mediate top-charm production at
interesting levels at muon\cite{ars} and $\gamma\gamma$\cite{hou} colliders.

Present constraints on the anomalous couplings in Eq. (\ref{effL}) arise 
from a 
variety of processes.  A global analysis of the flavor changing neutral
current processes 
$K_L\to \mu^+\mu^-\,, B\to \ell^+\ell^-X\,, K_L-K_S$ mass difference, and 
$B^0-\bar B^0$ mixing, as well as 
the oblique parameters in electroweak precision 
measurements has been performed\cite{hpz} for the dimension four operators
by forming a low-energy effective interaction after integrating out the 
heavy 
top-quark.  This procedure yields the restrictions
\begin{eqnarray}
g_L^Z=v_Z-a_Z &\leq& 0.05 \,,\nonumber\\
g_R^Z=v_Z+a_Z &\leq& 0.29  \,,
\end{eqnarray}
assuming a cutoff of 1 TeV.  Bounds on the $tc\gamma$ interactions can be
obtained from $B\to X_s\gamma$ and restrict\cite{hwyz} $|\kappa_\gamma|< 
0.1$,
using the normalization in (\ref{effL}) and assuming 
$\tilde\kappa_\gamma=0$.  
CDF has performed a direct search
for FCNC top decays and has placed\cite{cdf} the $95\%$ C.L. 
limits of $B(t\to q\gamma)<3.2\%$ and $B(t\to qZ)<33\%$, where $q=c$ or $u$.
In the photonic channel, this gives the constraint of 
$\kappa_\gamma\lsim 0.26$, which is not
yet competitive with the indirect bounds from $B\to X_s\gamma$.  The $Z$ 
decay channel is not yet at an interesting level of sensitivity.  These 
direct
constraints from top decays
are expected\cite{hpz,hwyz} to improve to the level of
$\kappa_\gamma\simeq 0.04$ and $\sqrt{v_Z^2+a_Z^2}\simeq 0.11$ during Run II
at the Tevatron with $10~\infb$ of integrated luminosity, and 
$\kappa_\gamma\simeq 0.0035$ and $\sqrt{v_Z^2+a_Z^2}\simeq 0.014$ at the LHC
with $100~\infb$ of integrated luminosity.  In addition, a $\sqrt s=400$ GeV 
photon collider can probe\cite{kerry} 
$\kappa_\gamma$ down to values of $\simeq 0.01$ with $10~\infb$ of 
integrated
luminosity via the reaction $\gamma\gamma\to t\bar q+\bar tq$.  Similar
studies have also been performed\cite{glue} for analogous anomalous
$tcg$ couplings.

The production rate for $\epem\to t\bar c+\bar tc$ was first
computed in Ref.~\cite{hikasa} with the dimention-four terms.
They have been later studied for the case with leptoquark
exchanges \cite{bh}, and for the case of flavor 
non-conserving Multi-Higgs-Doublet models in
Ref. \cite{atwood}, in Supersymmetry with R-parity violation 
\cite{rpar}, and for some models of mass matrix textures\cite{massmatrix}.
The SM one-loop induced production in \epem\ collisions was discussed
in \cite{chang} and other higher order processes,
such as $\epem\to t\bar c\nu_e \bar\nu_e$, 
have also been considered\cite{highord}.
It is the intention of this work to study the model independent case,
and to, more importantly, investigate the issue of detecting the signal
over the background and to determine to what precision the FCNC couplings
in Eq. (\ref{effL}) can be measured in \epem\ collisions.  We note that
the analysis presented here can also be applied to top-up-quark associated
production.

\section{Top-Charm Associated Production}

\subsection{The Total Cross Section}

\begin{figure}[t]
\centerline{
\psfig{figure=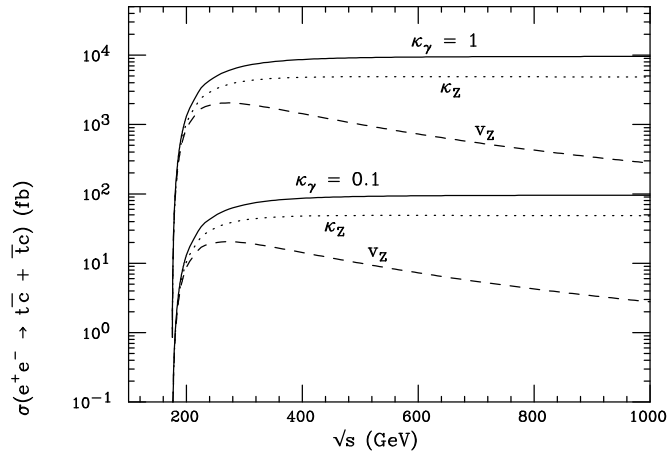,width=4in,angle=-90}}
\vspace*{-0.75cm}
\caption[]{Cross section for top-charm associated production as a function 
of
center-of-mass energy in \epem\ collisions.  Only one anomalous coupling
is taken to be non-zero at a time, with values as indicated.}
\label{xsec}
\end{figure}

We now investigate top-charm associated production in high energy \epem\
collisions.  This process is mediated by $s$-channel $\gamma^*,Z$ exchange,
\begin{equation}
\epem\to \gamma^*,Z\to t\bar c+\bar tc\,,
\label{signal}
\end{equation}
via the FCNC couplings.
Using the effective Lagrangian in Eq. (\ref{effL}) the differential cross 
section is calculated to be
\begin{eqnarray}
\label{xseceqn}
{d\sigma\over dz} & = & {3\pi\alpha^2\over 2s}\left( 1-{m_t^2\over 
s}\right)^2
\sum_{i,j=\gamma,Z} P_{ij} \left\{ B_{ij}
\left[\left( 1+{m_t^2\over s}\right)+\left( 1-{m_t^2\over 
s}\right)z^2\right]
\right.\nonumber\\
& &\left. +2C_{ij}z+{s\over m_t^2}D_{ij}
\left[\left( 1+{m_t^2\over s}\right)-\left( 1-{m_t^2\over 
s}\right)z^2\right]
+2E_{ij}+2F_{ij}z\right\}\,, 
\end{eqnarray}
where $z=\cos\theta$ with $\theta$ being the angle between the top-quark and 
the electron.  The usual propagator factor is defined as
\begin{equation}
P_{ij}={s^2[(s-M_i^2)(s-M_j^2)+(\Gamma_iM_i)(\Gamma_jM_j)]\over
[(s-M_i^2)^2+(\Gamma_iM_i)^2][(s-M_j^2)^2+(\Gamma_jM_j)^2]}\,,
\end{equation}
where $M_i\,,\Gamma_i$ refer the mass and width of the $ith$ gauge boson,
and the coupling factors are
\begin{eqnarray}
B_{ij} & = & (v_iv_j+a_ia_j)_e(v_iv_j+a_ia_j)_{tc}\,,\nonumber\\
C_{ij} & = & (v_ia_j+v_ja_i)_e(v_ia_j+v_ja_i)_{tc}\,,\nonumber\\
D_{ij} & = & (v_iv_j+a_ia_j)_e(\kappa_i\kappa_j+\tilde\kappa_i\tilde
\kappa_j)_{tc}\,,\\
E_{ij} & = & (v_iv_j+a_ia_j)_e(v_i\kappa_j+v_j\kappa_i)_{tc}\,,\nonumber\\
F_{ij} & = & (v_ia_j+v_ja_i)_e(a_i\kappa_j+a_j\kappa_i)_{tc}\,,\nonumber
\end{eqnarray}
with $v_\gamma=a_\gamma=0$.
The terms proportional to $C_{ij},F_{ij}$ are odd in $z$ and will produce an
asymmetric angular distribution if more than one anomalous coupling is
simultaneously non-vanishing.  
Due to the $\sigma_{\mu\nu}q^\nu$ structure of the dimension 
five operator, the term proportional to $D_{ij}$ does not have the usual
$1/s$ dependence and will dominate the cross section as the center-of-mass
energy increases.
Note that if only one anomalous coupling is non-zero at a time
(as we will assume implicitly unless stated otherwise), then
the integrated cross section is directly proportional to its square. 
Figure \ref{xsec} displays the total cross section
as a function of the center-of-mass energy, taking only one coupling
non-vanishing at a time with either $\kappa_\gamma=1(0.1)$,
$\kappa_Z=1(0.1)$, or $v_Z=1(0.1)$.  These values were chosen for purposes
of demonstration only, and the property that the cross section
is proportional to the square of the coupling is explicitly demonstrated.
From Eq. (\ref{xseceqn}) it is clear that taking $v_Z$ versus $a_Z$ (or
$\tilde\kappa_{\gamma,Z}$ versus $\kappa_{\gamma,Z}$) to be non-zero yields 
the 
same numerical result.

\vspace*{-0.5cm}
\nn

\subsection{Signal And Background}

We concentrate on the semi-leptonic decay of the single top-quark,
\begin{equation}
\epem\to \bar ct\to\bar cb\ \ell\nu_\ell,
\label{sign}
\end{equation}
to efficiently separate the signal from the SM background,
taking $\ell=e$ or $\mu$, and the charge-conjugate state is implied.
The irreducible SM background arises from $\epem\to W^+W^-\to 
\bar cb\ell\nu$ and is fortunately negligible due to the small size of the 
Cabbibo-Kobayashi-Maskawa mixing matrix element $V_{cb}$.  However,
without the ability to perfectly tag heavy flavor states, light quark jets 
are also a source of background.  The leading SM background then 
comes from the final state 
\begin{equation}
\epem\to \bar q q' \ell \nu,
\label{back}
\end{equation}
where $q,q'$ are light quarks.  This originates mainly from $W$ pair 
production 
as well as from $W$ bremsstrahlung in $\epem\to W+$2-jets. 
The total cross sections, including
the leptonic branching fractions and with no kinematical cuts, are presented 
in Table~\ref{nocuts} for the signal Eq.~(\ref{sign}) and the background 
Eq.~(\ref{back})  for three representative center-of-mass energies at 
$\epem$ colliders.
The signal cross sections are evaluated with one anomalous coupling
to be non-zero (equal to unity) at a time. The results for
$\kappa_\gamma, v_Z^{}$ and $\kappa_Z^{}$ are the same
as those for $\tilde\kappa_\gamma, a_Z^{}$ 
and $\tilde\kappa_Z^{}$, respectively.  We see that for these large values
of $\kappa_{\gamma,Z}$ the cross sections for
the dimension five operators are already competitive with
the background rates, even before any kinematical cuts are applied.

\begin{table}
\centering
\begin{tabular}{|c|c|c|c|} \hline\hline
  & 192 GeV  & 0.5 TeV  &  1 TeV \\ \hline
signal[$\kappa_\gamma=1$]  & 156 & 1980 & 2070 \\
signal[$v_Z=1$]    & 114 & 217 & 60 \\
signal[$\kappa_Z=1$]    & 130& 1060 & 1050 \\
bckgrnd & 5687 & 2252 & 864\\ \hline\hline
\end{tabular}
\caption[]{Total cross sections in fb for signal Eq.~(\ref{sign}) 
and background Eq.~(\ref{back}) including the leptonic branching 
fractions and with no kinematical cuts.}
\label{nocuts}
\end{table}

To roughly simulate the experimental environment, we first adopt the basic
kinematical cuts on the energy and pseudo-rapidity for the jets
and leptons
\begin{equation}
E_{j,\ell} > 10\ {\rm GeV},\quad  |\eta_{j,\ell}| < 2,
\label{cutsI}
\end{equation}
which corresponds to a 15-degree polar angle with respect to
the beam. We also smear the energies with a Gaussian standard
deviation of the detector response\cite{nld}
\begin{equation}
\Delta E/E = 40\%, \quad 10\%, 
\end{equation}
for jets and leptons, respectively.

\begin{figure}
\centerline{
\psfig{figure=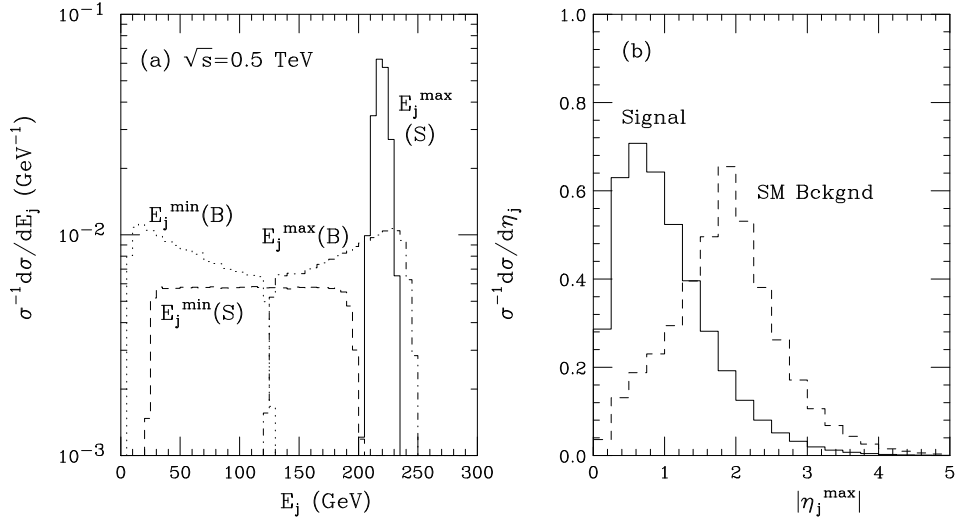,width=5in}}
\caption[]{Normalized jet-energy and rapidity distributions at 
$\sqrt s=500$ GeV.
(a) Jet energy distributions for the signal (S) and background (B) 
with the harder jet being labeled as $E_j^{max}$ and the softer jet 
as $E_j^{min}$.
(b) Jet rapidity distribution for the signal and the background.}
\label{jetdists}
\end{figure}

\begin{figure}
\centerline{
\psfig{figure=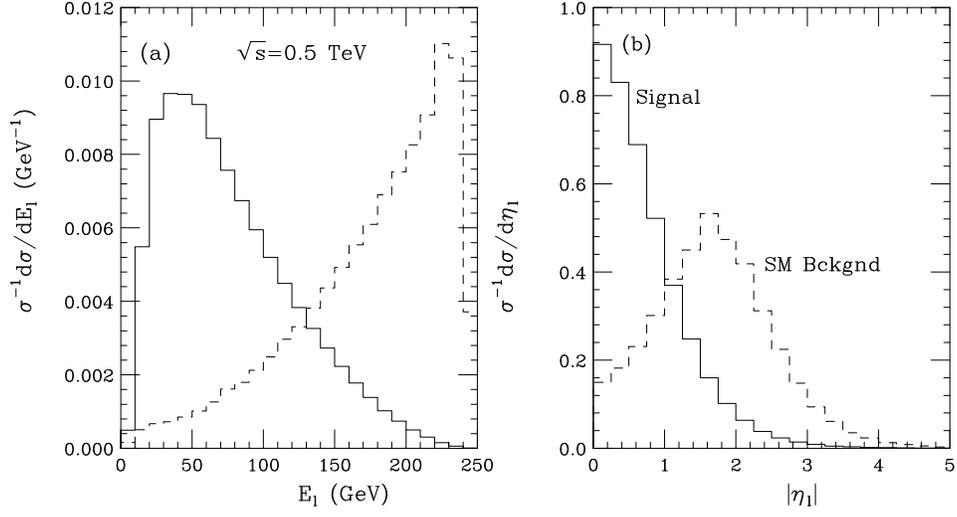,width=5in}}
\caption[]{(a) Normalized lepton-energy and (b) rapidity distributions
at $\sqrt s=500$ GeV for the signal (solid) and background (dashes).}
\label{leptdists}
\end{figure}

\begin{figure}
\centerline{
\psfig{figure=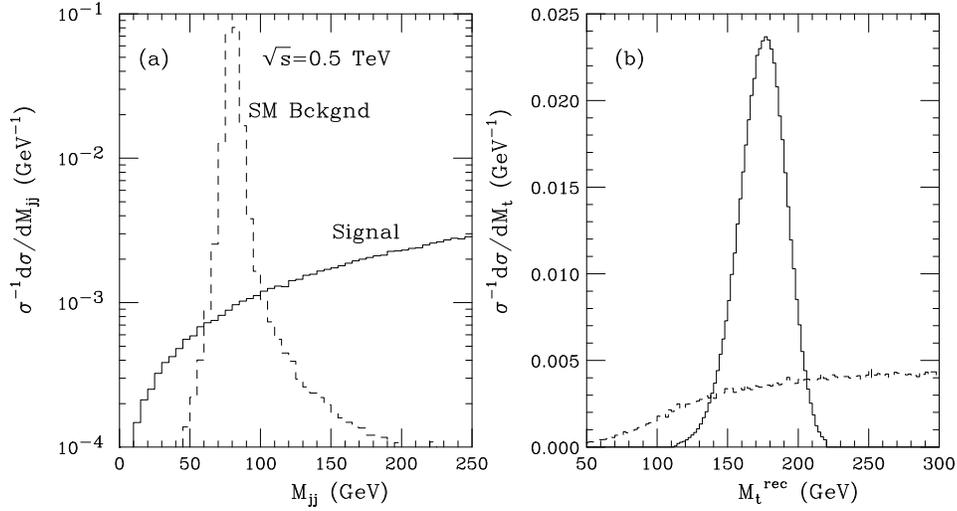,width=5in}}
\caption[]{Normalized distributions for the signal (solid) and background 
(dashes) at $\sqrt s=500$ GeV; (a) di-jet invariant mass distributions, and
(b) the reconstructed top-quark mass according to Eq.~(\ref{mtrec}).}
\label{recontop}
\end{figure}

Although the signal cross section is not expected to be very large
for values of the anomalous couplings which are consistent with
model expectations, 
the signal final state can be quite characteristic in comparison
with the background events.
First, due to the nature of two-body kinematics for the signal,
the charm-jet energy is fixed as 
\begin{equation}
E_c = {\sqrt s\over 2}(1-m_t^2/s),
\end{equation}
which leads to the values of $E_c\simeq 16\,, 220$, and 485 GeV at
$\sqrt s=192\,, 500$, and 1000 GeV, respectively. Similarly, the $b$-quark
energy from the top-quark decay in the signal is typically  
$E_b = {m_t/ 2} (1-M_W^2/m_t)\approx 70$ GeV,
modulo some smearing from the top-quark motion.
In addition, the signal jets are more central than those arising from 
the background sources in which there is a strong boost of
the $W$ system at higher energies. These features are displayed in
Fig.~\ref{jetdists} for $\sqrt s=500$ GeV. From Fig.~\ref{jetdists}(a), we 
see that in contrast to the nearly mono-value for the energy of the harder 
jet
in the case of the signal, the corresponding distribution for the background
is uniformly distributed.  Fig.~\ref{jetdists}(b) shows the 
difference in the rapidity distribution between the signal and the 
background.
Secondly, the charged lepton momentum for the SM background tends 
to be parallel to that of the parent $W^\pm$ boson, while the 
opposite holds for the signal. This is due to spin correlation
effects for transversely polarized $W$ bosons. Consequently, the 
charged lepton energy distribution for the background is harder 
due to the parallel boost by the $W$ system, while it is softer 
for the signal. This
is shown in Fig.~\ref{leptdists}, where (a) the $E_\ell$ spectra for the 
signal and the background are contrasted, and (b) 
the rapidity distributions are presented.
Another apparent difference between the signal and background kinematical
features is that the di-jet invariant mass for the background events 
primarily
reconstructs to $M_W$, as depicted in Fig.~\ref{recontop}(a). 
Finally, the most important confirmation of the signal is
the reconstruction of the top-quark mass. Although the missing energy from
the final state neutrino prevents a direct mass 
measurement of the top-quark, $m_t$ can 
still be accurately reconstructed from knowledge of the center-of-mass 
energy
and the charm jet energy via
\begin{equation}
m_t^{\rm rec} = (s- 2 {\sqrt s} E_c)^{1/2}.
\label{mtrec}
\end{equation}
This variable is depicted in Fig.~\ref{recontop}(b) for $\sqrt s=500$ GeV, 
where the discrimination power against the SM background is clearly 
observable.
The width of the $m_t^{\rm rec}$ distribution increases 
at higher energies due to the larger charm-jet 
energy smearing.

The above kinematical results presented in
Figs.~\ref{jetdists}-\ref{recontop} provide the motivation for the
optimization of our selective cuts.
These cuts are delineated in Table~\ref{cuts}. 
Employing these cuts, the cross sections are then calculated under the 
reconstructed top-quark mass
peak (as prescribed in the last column in Table~\ref{cuts}) with the results
being given in Table \ref{acuts}.  Here we see that our choice of 
kinematical
cuts are very efficient in reducing the size of the background while
retaining the signal.

\begin{table}
\centering
\begin{tabular}{|c|c|c|c|c|c|} \hline\hline
$\sqrt s$ & $E_j$(high) & $E_j$(low) & $|M_{jj} - M_W|$ & $E_\ell$ 
& $|m_t^{\rm rec}-m_t|$ \\ \hline
192 GeV & $> 60$  GeV & $< 20$ GeV & $>10$ GeV & -           & $<5$ GeV   \\
500 GeV & $> 200$ GeV & $> 20$ GeV & $>10$ GeV & $< 150$ GeV & $<40$ GeV  \\
1 TeV   & $> 460$ GeV & $> 20$ GeV & $>10$ GeV & $< 350$ GeV & $<100$ GeV \\ 
\hline\hline
\end{tabular}
\caption[]{Kinematical cuts for the event selection at different
collider energies.}
\label{cuts}
\end{table}

\begin{table}
\centering
\begin{tabular}{|c|c|c|c|} \hline\hline
    & 192 GeV & 0.5 TeV & 1 TeV \\ \hline
signal[$\kappa_\gamma=1$] & 129  & 1690  & 1880 \\
signal[$v_Z=1$]           &  95  &  169  &  49.7\\
signal[$\kappa_Z=1$]      & 108  & 900   & 951 \\
bckgrnd                   & 23.6 & 5.1  & 1.7 \\ \hline\hline
\end{tabular}
\caption[]{Total cross sections in fb for the signal Eq.~(\ref{sign}) 
and background Eq.~(\ref{back}) including the leptonic branching 
fractions and with the kinematical cuts presented in Table~\ref{cuts}.}
\label{acuts}
\end{table}

\subsection{Sensitivity to the anomalous couplings}

Given the efficient signal identification and substantial background 
suppression achieved in the previous section, 
we now estimate the sensitivity to the anomalous
couplings from this reaction using Gaussian statistics, which is applicable
for large event samples.  Here,
\begin{equation}
\sigma = {N_S\over \sqrt{N_S+B_S}},
\end{equation}
with $N_S$ and $N_B$ being the number of signal and background
events. We demand that $\sigma \ge 3$ in order to observe the signal, which 
approximately corresponds to the $95\%$
Confidence Level. 

\begin{figure}[h]
\centerline{
\psfig{figure=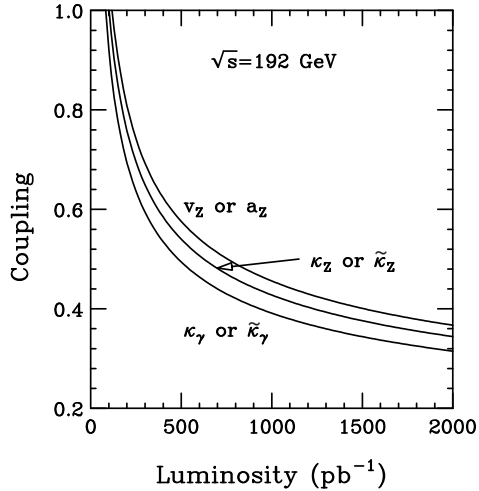,width=2.5in}}

\caption[]{$95\%$ C.L.
sensitivity to the FCNC couplings  at LEPII with $\sqrt s=192$ GeV
as a function of integrated
luminosity, summed over all four detectors. }
\label{lepii}
\end{figure}

At LEPII energies, the results in Table \ref{acuts} demonstrate that the 
$W^+W^-$ background is large while the
signal rate is relatively low. Figure~\ref{lepii} presents the
$95\%$ C.L. sensitivity to the FCNC couplings as a function of the 
integrated
luminosity, summed over all four detectors, with $\sqrt s=192$ GeV for 
$\kappa_\gamma (\tilde\kappa_\gamma)$, $v_Z^{}(a_Z^{})$,
and $\kappa_Z^{} (\tilde\kappa_Z^{})$. We see that the combined
sensitivity for 500 pb$^{-1}$ per detector could reach the 0.3$-$0.4 level.
This is similar to (or slightly worse than) 
the current indirect constraints obtained from 
rare decays.  Since the $b$-flavor tagging efficiency is
not very high at LEPII, being only $\sim 25-30\%$,
requiring a tagged $b$ in the final state actually
decreases the achievable sensitivity.

\begin{figure}
\centerline{
\psfig{figure=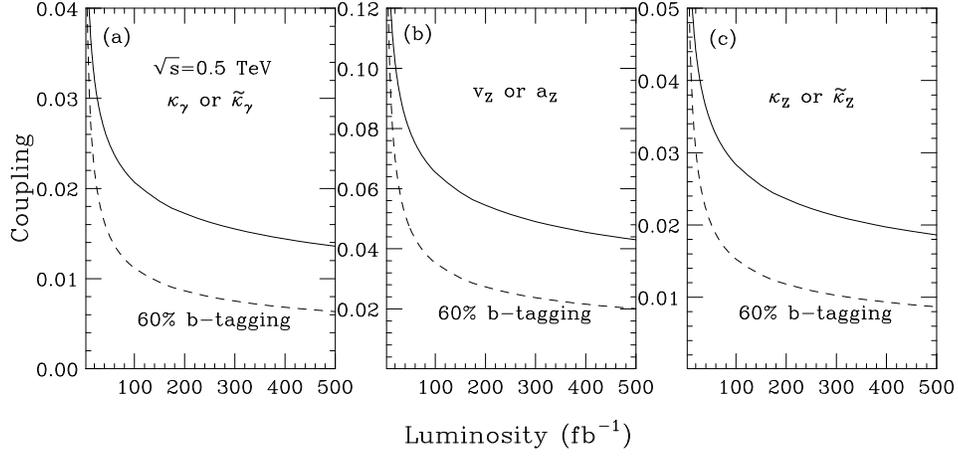,width=5in}}

\caption[]{$95\%$ C.L. sensitivity to the FCNC couplings as a function of
the integrated luminosity with 
$\sqrt s=500$ GeV for (a) $\kappa_\gamma$, (b) $v_Z$,
and (c) $\kappa_Z$, with and without b-tagging as labeled.}
\label{halftev}
\end{figure}

\begin{figure}
\centerline{
\psfig{figure=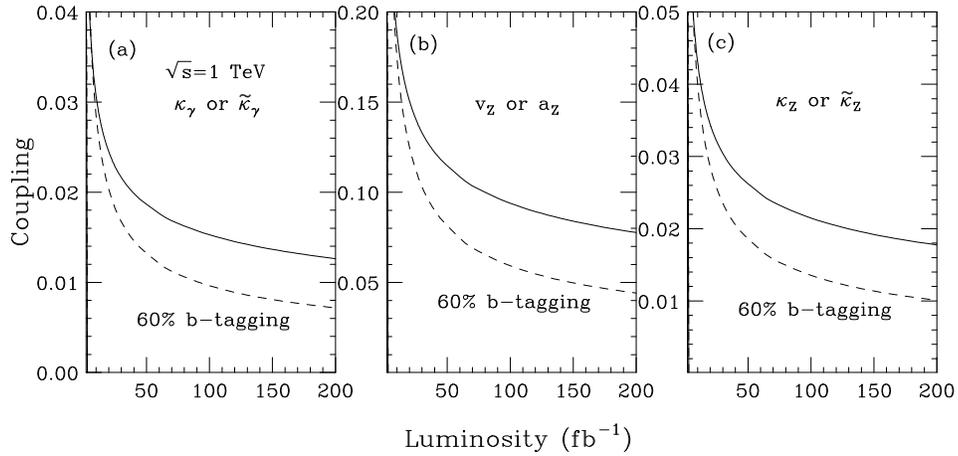,width=5in}}

\caption[]{$95\%$ C.L. sensitivity to the FCNC couplings as a function of
the integrated luminosity with $\sqrt s=1$ TeV for (a) $\kappa_\gamma$, 
(b) $v_Z$, and (c) $\kappa_Z$, with and without b-tagging as labeled.}
\label{onetev}
\end{figure}

At higher energies, the ability to probe the existence of
FCNC anomalous couplings is
greatly improved.  Figures~\ref{halftev} and \ref{onetev} display the
$95\%$ C.L. sensitivity to the couplings as a function of integrated
luminosity at $\sqrt s=0.5$ and 1 TeV for 
$\kappa_\gamma (\tilde\kappa_\gamma)$, $v_Z^{}(a_Z^{})$,
and $\kappa_Z^{} (\tilde\kappa_Z^{})$.  Here, the solid curves correspond to
the results using the kinematical cuts only, while the dashed curves include
the effects of b-tagging.  It is expected\cite{vxd4} that a CCD-based pixel
vertex detector combined with topological vertexing can achieve a 
$\sim 60\%$ b-quark identification efficiency with 
very high purity at high energy linear colliders.  This is not too far of an
extrapolation from the present $\sim 50\%$ b-quark identification efficiency 
that has recently been attained at SLD\cite{sal}.
We have extended the integrated luminosity for the 0.5 TeV linear collider
to 500 fb$^{-1}$, corresponding to expectations from the TESLA linear collider
design\cite{tesla}.  The sensitivity to these couplings
scales with the integrated luminosity ($L$) approximately as $1/\sqrt L$ 
when the
background is small. As a result, the ability to explore the FCNC couplings
is improved by more than a factor of 2 when the integrated
luminosity is increased from 50 to 500 fb$^{-1}$.  We also note that the
sensitivity to the couplings of the dimension five operators, 
$~^(\tilde\kappa^)_\gamma,~^(\tilde\kappa^)_Z$,  is increased by roughly
$30\%$ as the center-of-mass energy is raised
from 0.5 to 1 TeV.  This is as expected due to the structure of these 
operators.

\section{Asymmetries}

\begin{figure}[htbp]
\centerline{
\psfig{figure=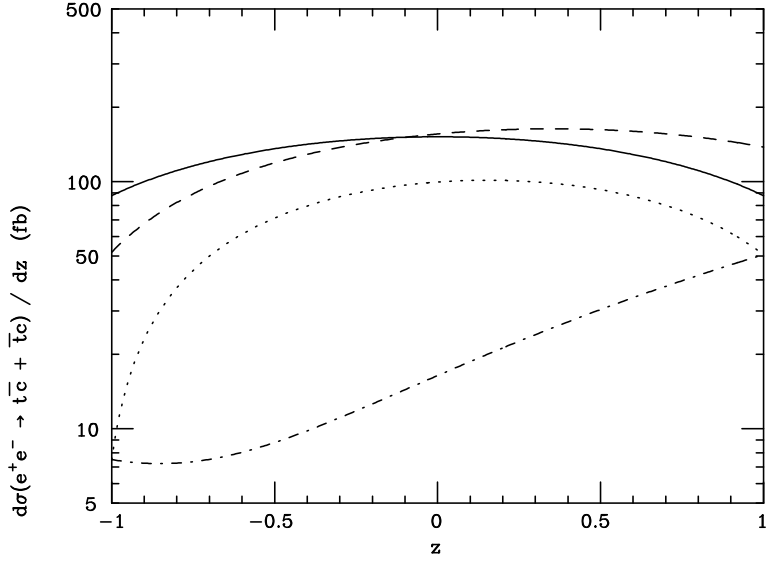,width=5in,angle=-90}}
\vspace*{-0.5cm}
\caption[]{Angular distributions for the
cases of $v_Z=a_Z=\kappa_Z=0.1$ (dashed curve), 
$a_Z=0\,, v_Z=\kappa_Z=0.1$ (solid), $v_Z=0\,, a_Z=\kappa_Z=0.1$
(dotted), and $v_Z=a_Z=0.1\,, \kappa_Z=0$ (dash-dotted) with
$\sqrt s=500$ GeV.}
\label{zdist}
\end{figure}

\begin{figure}[htbp]
\centerline{
\psfig{figure=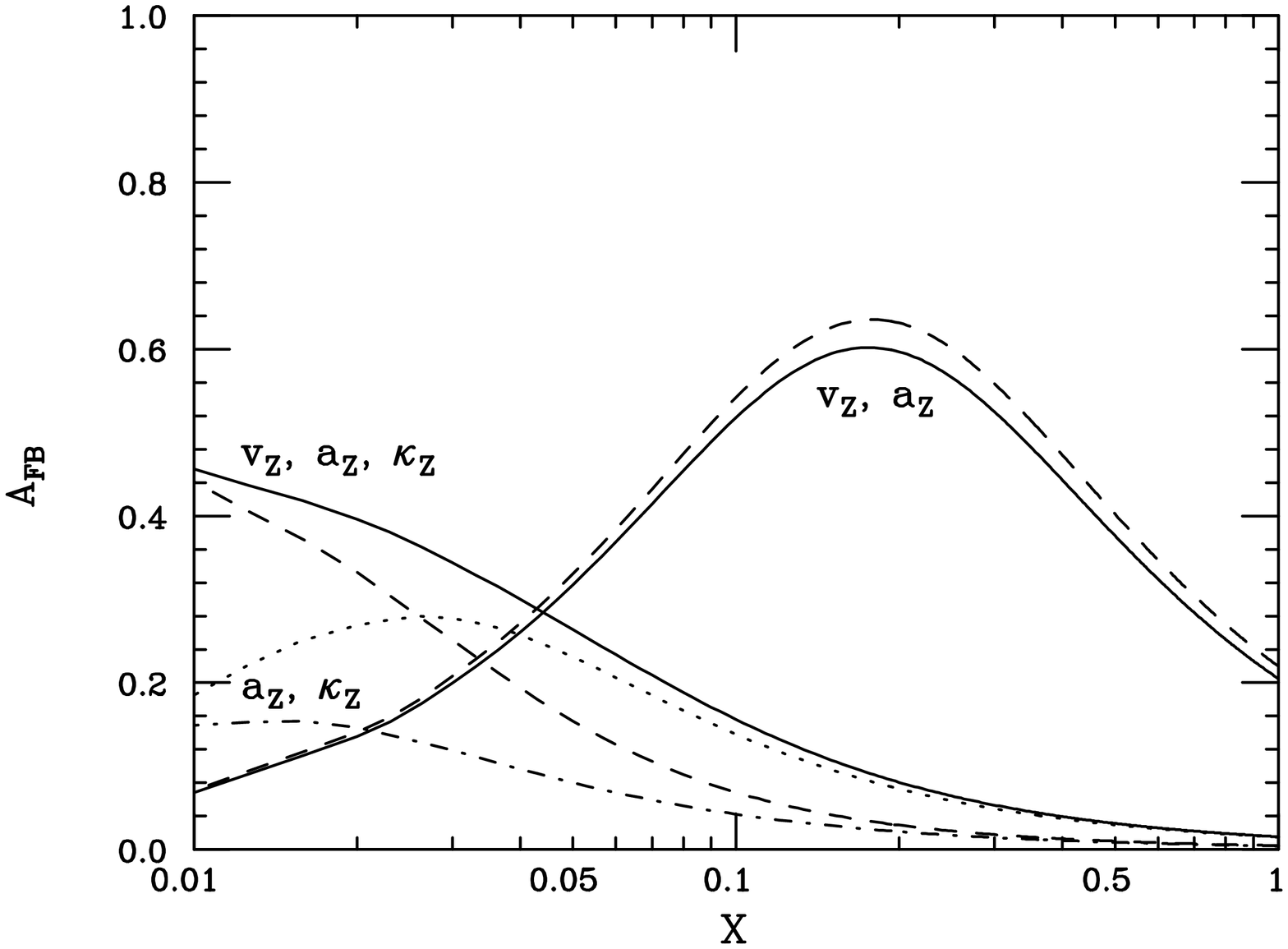,width=3.25in,angle=-90}
\hspace*{-5mm}
\psfig{figure=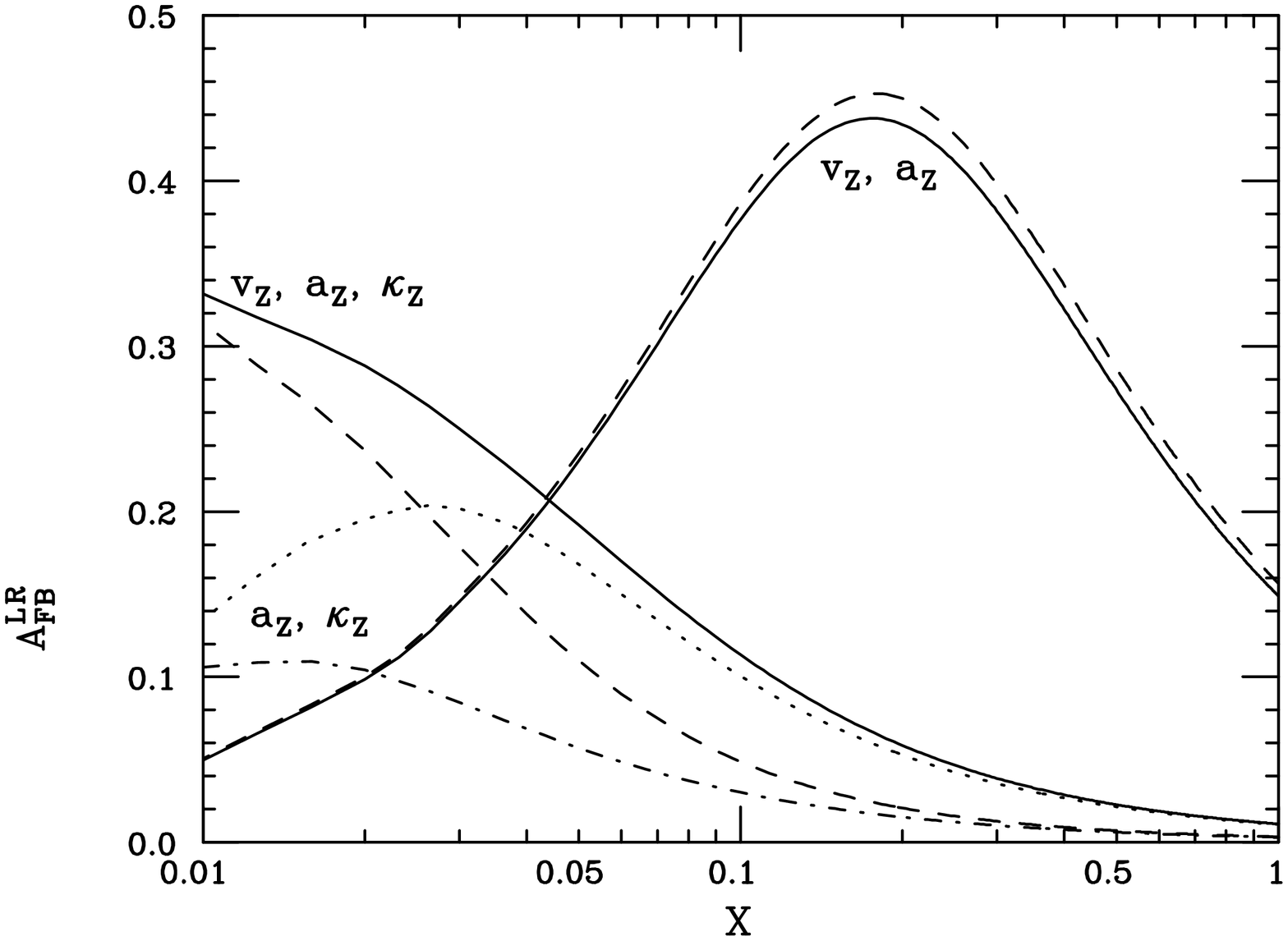,width=3.25in,angle=-90}}
\vspace*{-0.5cm}
\caption[]{(a) Forward-backward asymmetry and
(b) Left-Right forward-backward asymmetry as a function of the value of
the FCNC coupling.  The coupling combinations taken to be non-vanishing
and equal are as indicated.  For the $v_Z=a_Z$ and $v_Z=a_Z=\kappa_Z$ cases, 
the solid (dashed) curve corresponds to $\sqrt s= 0.5$ (1) TeV; the
$a_Z=\kappa_Z$ case is dotted (dash-dotted) for $\sqrt s=0.5$ (1) TeV.}
\label{asymm}
\end{figure}

As mentioned above, the odd terms in $\cos\theta$ in the cross section
generate an asymmetric angular distribution.  This is
illustrated in Fig. \ref{zdist} with $\sqrt s=500$ GeV for the sample 
cases of $v_Z=a_Z=\kappa_Z=0.1$ (dashed curve), 
$a_Z=0\,, v_Z=\kappa_Z=0.1$ (solid), $v_Z=0\,, a_Z=\kappa_Z=0.1$
(dotted), and $v_Z=a_Z=0.1\,, \kappa_Z=0$ (dash-dotted).  Note that
the case where $a_Z=0$ produces a symmetric distribution as expected from
the form of the cross section.  The other coupling combinations where $a_Z$ 
is non-vanishing all produce distinct asymmetric distributions which can
be used to distinguish between the various scenarios.  The resulting
forward-backward asymmetry and polarized left-right forward-backward
asymmetry are displayed in Fig. \ref{asymm}(a) and (b) as a function of the
value of the anomalous couplings for $\sqrt s=0.5$ and 1 TeV.  The coupling 
combinations taken to be non-zero with equal values are as indicated.  
The polarized asymmetry
is proportional to the FCNC coupling factors
\be
A_{FB}^{LR}\sim (v_iv_j+a_ia_j)_e(v_ia_j+v_ja_i)_{tc}+
(v_iv_j+a_ia_j)_e(\kappa_ia_j+\kappa_ja_i)_{tc}\,.
\ee
In Fig.~\ref{asymm} 
we have taken the degree of beam polarization to be $90\%$ and
employ a $10^\circ$ angular cut around the beam pipe to remove backgrounds
from the interaction region.  We see that these two asymmetries have
similar shapes, and hence the beam polarization does not add much new
information, however, they clearly distinguish between the various options for
the non-vanishing couplings.  Hence, if top-charm associated production is
observed, these asymmetries would provide a valuable tool for discerning
the structure of the FCNC couplings and unraveling the underlying physics.

\section{Conclusions}  

We have examined the possibility of top-charm associated production in
\epem\ collisions via FCNC couplings.  This mechanism, in contrast to the
study of rare top-quark decays, allows for the
exploration of higher dimensional operators at large values of momenta.
We used a model independent 
parameterization to describe these couplings and devised a set of cuts
to cleanly distinguish the signal from the background.  Our results
show that LEPII will be able to probe these couplings only at a level
which is comparable to the constraints from present data.  However, a
500 GeV linear collider with 50~\infb\ has more sensitivity to these
couplings than the Tevatron with 30 fb$^{-1}$, and with 500 fb$^{-1}$ it is
comparable in reach to that of the LHC with 100 fb$^{-1}$.  The 1 TeV
machine gives roughly a $30\%$ improvement in sensitivity.  These 
cases correspond to exploring FCNC top decays with branching ratios in
the range $10^{-4}-10^{-3}$.  In addition, if $a_Z\ne 0$, then angular
and polarization asymmetries can be formed which can yield information
on the structure of the couplings and the underlying physics.
\vspace{0.5cm}

\noindent {\bf Acknowledgments}:
We would like to thank G. Burdman, J. Jaros, F. Paige, M. Peskin, 
K. Riles and T. Rizzo for discussions related to this work.
T.H. was supported in part by a DOE grant No. DE-FG02-95ER40896 
and in part by the Wisconsin Alumni Research Foundation.

%
\def\IJMP #1 #2 #3 {Int. J. Mod. Phys. {\bf#1},\ #2 (#3)}
\def\MPL #1 #2 #3 {Mod. Phys. Lett. {\bf#1},\ #2 (#3)}
\def\NPB #1 #2 #3 {Nucl. Phys. {\bf#1},\ #2 (#3)}
\def\PLBold #1 #2 #3 {Phys. Lett. {\bf#1},\ #2 (#3)}
\def\PLB #1 #2 #3 {Phys. Lett. {\bf#1},\ #2 (#3)}
\def\PR #1 #2 #3 {Phys. Rep. {\bf#1},\ #2 (#3)}
\def\PRD #1 #2 #3 {Phys. Rev. {\bf#1},\ #2 (#3)}
\def\PRL #1 #2 #3 {Phys. Rev. Lett. {\bf#1},\ #2 (#3)}
\def\PTT #1 #2 #3 {Prog. Theor. Phys. {\bf#1},\ #2 (#3)}
\def\RMP #1 #2 #3 {Rev. Mod. Phys. {\bf#1},\ #2 (#3)}
\def\ZPC #1 #2 #3 {Z. Phys. {\bf#1},\ #2 (#3)}

\end{document}